\documentclass[aps, prd, twocolumn, lengthcheck, superscriptaddress, showpacs, letterpaper, nofootinbib]{revtex4-1}
\usepackage{amsfonts}
\usepackage{amsmath}
\usepackage{amssymb}
\usepackage{xcolor}
\usepackage{graphicx}
\usepackage{subfigure}
\usepackage{longtable}
\usepackage{slashed}
\usepackage{hyperref}
\usepackage{amsthm}
\usepackage{mathrsfs}
\usepackage{color}
\usepackage{epsfig}
\usepackage{epstopdf}
\def\skyrme-model{skyrmion$_\pi$}

\def\be{\begin{eqnarray}}\def\ee{\end{eqnarray}}
\def\bi{\bibitem}\def\del{\partial}
\def\la{\langle}\def\ra{\rangle}
\def\be{\begin{eqnarray}}\def\ee{\end{eqnarray}}
\def\lsim{\mathrel{\rlap{\lower3pt\hbox{\hskip1pt$\sim$}}
     \raise1pt\hbox{$<$}}} 
\def\gsim{\mathrel{\rlap{\lower3pt\hbox{\hskip1pt$\sim$}}
     \raise1pt\hbox{$>$}}} 

\allowdisplaybreaks

\begin{document}

\title{Mapping topology of  skyrmions and fractional quantum Hall droplets
\\  to nuclear EFT for ultra-dense baryonic matter}


\author{Mannque Rho}
\email{mannque.rho@ipht.fr}
\affiliation{Institut de Physique Th\'eorique, Universit\'e Paris-Saclay, CNRS, CEA, \\  91191 Gif-sur-Yvette c\'edex, France }

\date{\today}

\begin{abstract}
We describe the mapping at high density of topological structure of baryonic matter  to a nuclear effective field theory that implements hidden symmetries emergent from strong nuclear correlations. The theory so constructed is found to be consistent with no conflicts with the presently available observations in both normal nuclear matter and compact-star matter. The hidden symmetries involved are  ``local flavor symmetry" of the vector mesons identified to be (Seiberg-)dual to the gluons of QCD and hidden ``quantum scale symmetry" with an IR fixed point with a ``genuine dilaton (GD)" characterized by non-vanishing pion and dilaton decay constants.  Both the skyrmion topology for $N_f \geq 2$ baryons and the fractional quantum Hall (FQH) droplet topology for $N_f=1$ baryons are unified in the ``homogeneous/hidden" Wess-Zumino term in the hidden local symmetry (HLS) Lagrangian. The possible indispensable role of the FQH droplets in going beyond the density regime of compact stars approaching scale-chiral restoration is explored by moving toward the limit where both the dilaton and the pion go massless.
   
\end{abstract}
\maketitle
\setcounter{footnote}{0}

\section{Introduction}

\subsubsection*{The Problem}

In nuclear processes going from low density near normal nuclear matter to high density relevant to  massive compact stars, two or possibly more density regimes are  involved. They are most likely delineated by changes of degrees of freedom (DoFs). At low densities up to slightly above the equilibrium nuclear matter density ($n_0\simeq 0.16$ fm$^{-3}$), say, to $n\lsim 2 n_0$, the relevant degrees of freedom are the nucleons $N$ and  pions $\pi$ figuring in what is referred to as ``standard chiral effective field theory" (given the acronym $s$ChEFT) with a cut-off scale set, typically, at $\Lambda_{sChEFT}\sim (400-500)$ MeV. When treated to N$^m$LO for $m=(3-4)$ in systematic chiral power expansion,  {\it ab initio}  calculations in $s$ChEFT have been seen to work highly satisfactorily for nuclear structure in finite nuclei as well as for properties of nuclear matter. This impressive success in nuclear physics could be taken as a convincing proof of Weinberg's  ``Folk Theorem  for EFT" as applied to QCD~\cite{FT}\footnote{There are a large number of excellent reviews in the literature, much too numerous, however, to make an adequate referencing. We apologize for not listing them here.}. This $s$ChEFT is expected to be extendable to  $\sim 2$ times $n_0$, but it is to break down by the premise of  EFT at some high density as the relevant degrees of freedom are no longer just the nucleons and pions but more massive hadronic degrees of freedom, and ultimately the QCD degrees of freedom, i.e., quarks and gluons, must figure as density increases. Thus there must be one or more changes of DoFs from hadronic to quarkonic. 

Due to the paucity of trustful theoretical tools for guidance in the absence of lattice approach to QCD in dense medium --   in contrast to thermal matter --   there is no clear indication how many and in what form(s)  these changeovers of DoFs  could take place,  in the density regime relevant to the center of massive compact stars,  $\sim (5 - 7) n_0$. This presents a totally uncharted domain that could very well encompass several different fields, say, condensed matter, nuclear and particle in addition to astrophysics,

There are several treatments in the literature that invoke quarks in various different forms, perturbative or nonperturbative, to explore this uncharted domain. In this note we describe a possible strategy that exploits the topological structure of baryonic matter -- without explicit QCD variables -- as density increases beyond $\sim 2n_0$ to access the putative baryon-quark continuity. In this approach, no explicit quark-gluon DoFs are invoked, but fractional (baryon-charged) quasiparticles induced from topology change(s) seem to figure. 

Among various topological structures of baryons, we adopt the skyrmion topology for the number of flavor $N_f \geq 2$ and the fractional quantum Hall (FQH) topology for $N_f=1$. The former is intricately relevant to QCD -- i.e., large $N_c$ limit -- in nuclear physics at low density involving the u(p) and d(own) quarks and the latter will be relevant at some high density.\footnote{The s(trange) quark does enter essentially in the formulation both in the way the scalar dilaton and $\eta^\prime$ figure but we will focus first on the $N_f=2$ systems and then come to the case where $\eta^\prime$ could enter in FQH topology.} 

Ultimately this feat should be feasible in terms of skyrmions -- ignoring FQH topology -- with a Lagrangian modeling QCD with all relevant degrees of freedom and work out the full topological structure of baryons going beyond the Skyrme model~\cite{Skyrme} with pions only (denoted as \skyrme-model), but this is currently out of reach.  There is an on-going effort with the \skyrme-model at suitable chiral derivative orders~\cite{adam}, but it is mathematically daunting to do realistic calculations  even with the truncated model to confront Nature. Our strategy is to map what are considered -- though not firmly confirmed -- as {\it robust} topological inputs, independent of the details of the Lagrangian, to an EFT going beyond  $s$ChEFT -- that we shall give the acronym G$n$EFT standing for ``generalized nuclear EFT" --  and treat it in terms of  a Landau Fermi-liquid approach to baryonic matter which we put in the class of the density functional (``DF")  \`a la Hohenberg-Kohn theorem. 

This type of approach is highly unconventional,  admittedly incomplete in various aspects.  Nonetheless it is found not only to successfully post-dict the overall properties of nuclear matter at density $\sim n_0$ in quality more or less comparable to high chiral-order ($N\geq 3$) $s$ChEFT but also account for the properties of massive compact stars at $n \gg n_0$ in fair consistency with what has been established in astrophysical observations.  There is however a surprising new  prediction -- which is extremely simple -- that follows from the topology-change exploited in the approach, namely,  the ``pseudo-conformal" sound velocity of the star and its impact on the structure of the core of compact stars. Our claim is that it exposes certain emergent symmetries hidden in QCD at low density.
\subsubsection*{The Motivation}

The basic idea figuring here was largely inspired and motivated by the predominant role topology plays in quantum critical phenomena in condensed matter systems,  stunningly exemplified, among others, by the fractional quantized Hall (FQH) effects~\cite{tong}.  This idea is germinated by the observation that many-body interactions in strongly correlated condensed matter systems and in nuclear many-body systems, although the basic interactions are different, i.e., QED vs QCD,  intricately  share certain common features. In the former, there have been remarkable breakthroughs by formulating correlated electron problems in terms of topology, thereby mapping many-electron interactions to topological field theories and also the other way around. The strategy we  describe here is prompted to do something analogous to what has been done in the physics of  the quantum Hall effects (QHEs), with certain arguments -- and  ideas -- borrowed therefrom. This effort is currently largely unrecognized in the nuclear community.  Of course they are necessarily of different nature given that we are dealing with strong interactions (QCD) with inherently  more complex dynamics than in QED. Furthermore strong interactions have much less direct access to experiments than in condensed matter systems.
\subsubsection*{The Objective}
Let us first state briefly what the problem is and what the objective of this article is. 

The main stream of current activities in nuclear/astro-physics community motivated by what's heralded as ``first-principles" approaches to nuclear physics {\it and} gravity-wave signals of merging neutron stars is anchored on  {\it ab initio} treatments of the $s$ChEFT at higher chiral orders and its related density-functional-type theories {\it defined and valid} at the density regime $\lsim 2n_0$. They are then extrapolated, resorting to various sophisticated ``meta-modelings,"  relying on the Baysian inference, high-order ``uncertainty  analysis" etc., to higher densities relevant for massive compact stars. In doing this, one confronts the inherent obstacles due to the paucity of trustful theoretical tools, given the inaccessibility of lattice QCD to high density.

The spirit of this article is drastically different from the majority of approaches pursued in the field. 

The aim of the approach adopted in this note is instead to construct as simple and economical an effective field theory as possible,  implementing what are deemed to be necessary to meet the requirement for the ``Folk Theorem for EFT"  appropriate for normal as well as compact-star matter. The philosophy is then to see how far one can go forward with this extremely simple theory -- never mind the nitty-gritty ``error uncertainties" -- before being hit by ``torpedoes."\footnote{In the spirit of Farragut's famous uttering `` Damn the torpedoes. Full speed ahead!"} 

In a nutshell, this article attempts to explain why the approach developed by us -- which is admittedly (over)simplified and more intuitive than of rigor -- seems to work surprisingly well for nuclear dynamics ranging from low density at $n\sim n_0$ where the $s$ChEFT is {\it believed} to be applicable to high density $n\sim (5-7)n_0$ where it is {\it suspected} to break down. The objective of this article is to see to what extent we can offer justifications and what need to be improved upon for the results obtained up to date.

Although the approach presented here predates the arrival of $s$ChEFT in 1990's, the core idea was the principal theme of the 5 year ``World Class University Project" (WCUP)  established at Hanyang University in Seoul in 2007 funded by  the Korean Government. This WCUP was in some sense in anticipation of the upcoming ambitious Institute of Basic Science (IBS) with the purpose to put Korean basic science on the world's frontier. We will base our discussion on what was initiated in 2007 and continued after the termination of the WCUP/Hanyang up to today.

The results that we will refer to are mostly available in the literature.  Their up-to-date status will be discussed in an accompanying contribution by Yong-Liang Ma~\cite{YLM} to which we will refer for quantitative details. The development up to 2018 was summarized in \cite{WS-RM-2018}, written in tribute to Gerry Brown who had made invaluable contributions to the development of theoretical nuclear physics in Korea. More recent developments, which make the story more exciting,  are found in \cite{MR-PPNP} and \cite{MR-manifestation}.  We must say that this development remains more or less unrecognized in the field. We hope that this note makes the basic idea involved better understood.
\subsubsection*{The Strategy}\label{KS}
The best way to motivate the reliance on topology to go from nucleons and pions at low density to quarks (and gluons) at high density is to think of a possible parallel of the approach adopted to how the physics of quantum Hall effects  is formulated in terms of a topological field theory.  What we have in mind in particular is the mapping of the fractional quantum Hall effect (FQHE) given in Chern-Simons topological field theory  to the Kohn-Sham density functional theory (DFT)~\cite{DF-jain,jain}.   The parallel is of course far from direct, given the totally different physics involved, but what figures in  both the fractional quantum Hall effect and compact-star matter involves mapping between the {\it microscopic description}, DFT, and the {\it macroscopic description}, Chern-Simons field theory.  The possible presence of such a parallel, although present from the very beginning of the WCU/Hanyang program,  was only very recently recognized by us thanks to the on-going works of string-theory-oriented theorists. It is suggested that both the KS-FDT-type microscopic approach and the Chern-Simons field theory-type macroscopic approach figured conceptually in the development made at the WCU/Hanyang and since then. 

Briefly the parallel that we see is as follows. 

In \cite{DF-jain,jain}, the system of strongly interacting electrons in the FQHE regime is formulated in terms of composite fermions of electrons bound with even number of quantum vortices involving an $U(1)$ gauge field {\it emergent} from strong correlations of the electrons. The complex effects of many-electron interactions are cast in a single-particle formalism in Kohn-Sham (KS) density functional theory incorporating the emergent $U(1)$ gauge interactions  between {\it weakly interacting}  composite fermions (CFs), i.e., quasiparticles,  induced by the quantum mechanical vortices. The gist of the approach then is that {\it the topology of Chern-Simons field theory is translated into the effective field theory,  DFT}.  What's remarkable in this approach is that their DFT does ``faithfully capture the topological characteristics" of the FQHE.

The approach that we will follow is inspired from the analogy of accessing the strongly correlated strong interactions in the density regime $n\gsim (2-4)n_0$ -- where $s$ChEFT is presumably broken down -- to the mapping  of the FQHE to the problem of nearly non-interacting composite electrons in the KS-DFT subject to an emerging ``magnetic field."  This analogy is not totally unfamiliar in its generic form in nuclear dynamics where the effective field theories of QCD, e.g., $s$ChEFT at low density, are extended to higher density regime with the cutoff set higher than $\Lambda_{sChEFT}$ in the form of ``relativistic mean-field theory" with heavier meson DoFs included. In fact one could consider these EFT approaches -- including the G$n$EFT discussed in this note  -- to generically belong to the class of Kohn's DFTs (including KS-DFT) applied in nuclear physics.\footnote{There is in the literature a huge variety of ``mean-field theories," both relativistic and non-relativistic, for treating both finite nuclei and infinite nuclear matter.  Some of them are capable of explaining the ensemble of  available terrestrial and astrophysical observables with success. They remain however mostly phenomenological, having little if any to do with the fundamental theory QCD, at the densities relevant to massive compact stars.}

We define our principal strategy as follows: Incorporate into an EFT -- called  ``G$n$EFT" from here on  to be distinguished from the standard ChEFT --  what are deemed to be ``robust" properties of the dense skyrmion matter built with the DoFs heavier than the pion. This EFT  is to capture as ``faithfully" as feasible the topological characteristics of the skyrmion matter. The basic assumption made is that at high density and in the large $N_c$ limit, the skyrmion matter is a crystal with a {\it negligible} contribution from the kinetic energy term~\cite{crystal}.
 
We mention in anticipation that the $\omega$ meson figuring as the $U(1)$ component of the hidden local symmetry (HLS)  can be identified with the $U(1)$ Chern-Simons field playing a role in the fractional quantum Hall droplet structure for the $N_f=1$ baryon associated with the $\eta^\prime$ singularity. We will return to the $\eta^\prime$ singularity because it is currently argued to be crucial for chiral restoration involving topology at high density, which is most likely outside of the range of densities relevant to compact stars, an issue that has only very recently been raised. 

In short, what comes out from the approach treated in this note, simple  in concept  -- and  albeit unorthodox -- turns out to work well ranging from nuclear matter density to high density relevant to massive compact stars.  Up to date we see no serious tension with Nature as reviewed in \cite{YLM} for this Special Issue of MDPI.

\section{Topology in baryonic matter}
\subsection{Change of DoFs: Hidden Symmetries}
Limiting our considerations for the moment to the density regime relevant to stars of mass $\sim 2 M_\odot$, we will ignore the role that the FQH droplets may play in strong correlations involved. We will return to that matter later. To simulate the change of DoFs in terms of topology as density goes up from below to above the putative baryon-quark continuity density  denoted as $n_{\rm BQC}$, two hidden symmetries  invisible in QCD in the vacuum are found to be {\it absolutely} essential.   One is hidden local symmetry (HLS)~\cite{HLS} and the other is hidden scale symmetry (HSS). The cutoff scale involved for the G$n$EFT should be greater than the cutoff effective for $s$ChEFT. The precise value of the relevant cutoff scale is not needed for what follows, but to be specific, one can take the HLS scale to that given by the vector meson ($V=(\rho, \omega)$) mass $m_V\sim 700$ MeV. The  scale symmetry is associated with the possible dilaton scalar $f_0(500)$ which will be later considered as a pseudo-Nambu-Goldstone boson of broken scale symmetry.  

As for the baryon-quark crossover density, we will be considering $n_{\rm BQC}\approx (2-4)n_0$ which will be identified later with the topology change density $n_{1/2}$. 

Defining precisely what the hidden symmetries to be incorporated are requires the details which have been given in review articles, e.g., \cite{MR-PPNP}. The basic ideas can however be explained rather simply without losing physical content. Here we will summarize the key ingredients that enter in the G$n$EFT.

One of the two symmetries that play a crucial role in the G$n$EFT is the hidden local symmetry (HLS) first formulated in \cite{HLS} and made powerfully applicable to nuclear physics as comprehensively reviewed in \cite{HY:PR}.   An important property of the HLS  concerned is that it is a gauge symmetry {\it dynamically generated} giving rise to  ``composite" gauge field of pions. The existence of such a composite gauge boson is proven to be ``inevitable" if such  a symmetry is implicit in the dynamics~\cite{suzuki} as assumed in our approach. It implies that the vacuum could be tweaked under  extreme conditions, say, by high temperature or high density, such that 
\be
m_V\propto  g_V \to 0 \label{vm}
\ee
where $g_V$ is the gauge coupling for the vector mesons $V$.  The point (\ref{vm}) is  called  the  ``vector manifestation (VM) fixed point"~\cite{HY:PR}.  How to expose such a symmetry at high temperature as in heavy-ion dilepton experiments or at high density as in compact stars is an extremely subtle issue. An important point to note in this connection  is the modern realization that there is a possible duality \`a la Seiberg (referred to as ``Seiberg-(type)duality") between the vector mesons of HLS and the gluons of QCD. This involves a conjecture, but our study points to its validity as we will try to argue.

The other hidden symmetry that figures equally importantly is the scale symmetry. There is a long history with a still on-going controversy on how the scale symmetry is manifested in gauge theories, e.g., in strong-interaction physics, and in going beyond the Standard Model (BSM).  In the literature are found strong arguments that  ``dilatons do not exist in QCD for $N_f\sim 3$."  We differ from such arguments and eschew going into that highly controversial issue for which we refer to \cite{MR-PPNP,MR-manifestation} viewed vis-\`a-vis with nuclear physics. As argued there, what is relevant to G$n$EFT is the ``genuine dilaton (GD)" scenario of \cite{GDS}.\footnote{While this article was being drafted, there appeared an article in which a very similar IR fixed-point structure with a ``conformal dilaton" was arrived at~\cite{DDZ}. In our view, at some high density, the genuine dilaton will coincide with the conformal dilaton when the IR fixed point is approached. This matter will be discussed elsewhere.}  The GD senario posits that there is an infrared (IR) fixed point with $\beta (\alpha_{s\rm IR})=0$ in the chiral limit (with u(p), d(own) and s(trange) quark masses equal to 0) and that the $f_0(500)$ is the scalar pseudo-Nambu-Goldstone (pNG)  boson of spontaneously broken scale symmetry which is also explicitly broken by quantum (scale) anomaly. One of the most distinctive characteristics of this scenario is that the IR fixed point, which is most likely non-perturbative, is realized in the {\it NG mode} with non-zero dilaton condensate (or decay constant) and non-zero pion condensate (or decay constant)\footnote{We note in this footnote that this scenario differs basically from the scenario popular in the BSM circle working with $N_f\sim 8$ in the conformal window of the IR fixed point realized in the Wigner mode.}. 
Given that the dilaton $\chi$ is a pNG boson as the pions $\pi$ are and both satisfy soft theorems, one can make a systematic power counting expansion in chiral-scale-symmetric theory as in chiral-symmetric theory. The power counting in chiral expansion is well established. Scale symmetry brings an additional power counting in terms of the expansion of the $\beta$ function. Expanding the gluon stress-tensor $\beta$ function in the QCD coupling $\alpha_s$ near the IR fixed point,
\be
\beta(\alpha_s)=\epsilon \beta^\prime (\alpha_{s\rm IR})+ O(\epsilon^2)
\ee
where $\epsilon=\alpha_s -\alpha_{s\rm IR}$ and $\beta^\prime >0$.  The non-zero mass of $f_0$ is attributed to $ |\beta^\prime (\alpha_{sIR})\epsilon|$. Thus the power counting in the scale symmetry is
\be
\epsilon\sim O(p^2)\sim O(\del^2).
\ee
One can in principle make a systematic chiral-scale counting comparable to chiral symmetry~\cite{GDS,LMR}.   

The matter-free-space mass of the dilaton $\sim 500$ MeV is comparable to that of kaons, so the dilaton is put on the same mass scale as the octet pions. In principle, one has an $SU(3)$ chiral Lagrangian coupled to the dilaton. In the background of nuclear medium, however, as is well known from nuclear phenomena, the lowest-mass meson of the scalar quantum number  is expected to undergo significant  mass-drop whereas kaons do not appreciably. Therefore in medium, identifying the scalar meson to be the dilaton, one can ignore the s quark in chiral-scale dynamics at high density.  Unless otherwise stated, this will be what we will do. 

We write  the dilaton field as
\be
\chi=f_\chi e^{\sigma/f_\chi}
\ee which is referred, in the literature, to as ``conformal compensator" field that figures in the notion of ``quantum scale invariance (QSI)."  As written, $\chi$ transforms linearly under scale transformation whereas $\sigma$ transforms nonlinearly. One can use either. In what follows,  we find it more convenient to employ the linearly transforming field $\chi$.  The Lagrangian that combines the (octet) pion field $\pi$, the HLS fields $V=(\rho,\omega)$ and the dilaton field $\chi$, suitably written in a chiral-scale invariant way with the scale-symmetry breaking term put in the dilaton potential $V(\chi)$,  will be denoted ${\cal L}_{\chi {\rm HLS}}$ with $\chi$ standing for the dilaton. It is of the form
\begin{eqnarray}
{\cal L}_{\chi\rm HLS} & = & \left(\frac{\chi}{f_\chi}\right)^2 \left(f_\pi^2 {\rm Tr}\left[\hat{\alpha}_{\perp \mu}\hat{\alpha}_{\perp}^\mu\right] + a f_\pi^2 {\rm Tr}\left[\hat{\alpha}_{\parallel \mu}\hat{\alpha}_{\parallel}^\mu\right]\right) \nonumber\\ 
-&& \frac{1}{2g^2}{\rm Tr}\left[V_{\mu\nu}V^{\mu\nu}\right] +\cdots\nonumber\\
 +&& {\cal L}_{\rm hWZ}
 + \frac{1}{2} \partial_\mu \chi \partial^\mu \chi +V(\chi)
\label{MHLS}
\end{eqnarray}
where $\hat{\alpha}_{\parallel \mu}$ and  $\hat{\alpha}_{\perp \mu}$ are Maurer-Cartan 1-forms and $V(\chi)$ is the dilaton potential. ${\cal L}_{\rm hWZ}$ is what's known as ``homogeneous" (or ``hidden") Wess-Zumino term which is scale-invariant that will figure later. For $N_f >2$, there is the 5-D topological Wess-Zumino term which we have left  out as it does not figure in our discussions.


%
\subsection{Topology Change}\label{TC}
There is a growing evidence that skyrmions as nucleons could describe finite nuclei as well as infinite nuclear matter~\cite{multifacet},  but at present it is far from feasible to address dense compact star matter quantitatively in terms of the pure skyrmion structure. It is however found feasible to extract topological properties of dense baryonic matter by simulating skyrmions on crystal lattice. In doing this, it is assumed that the topological characteristics extracted from the skyrmion crystal can be taken as {\it robust} and be exploited for {\it making} the mapping of topology to density-functional (DF) theory. Of course considering skyrmions on the crystal  cannot be a good approximation for low-density matter.  Clearly  it would make little sense to think of nuclear matter, which is best described as a Fermi liquid, as a crystalline. However at high density and in the large $N_c$ limit, baryonic matter could very well be in the form of a crystal~\cite{crystal,dyonicsalt}. It appears quite reasonable -- and it is assumed in this note -- that the baryonic matter at a density greater than  the putative baryon-quark transition density denoted as $n_{BQC}$ could encapsulate certain characteristic features of topology that are {\it not} captured in $s$ChEFT-type approaches.

Consider skyrmions constructed with ${\cal L}_{\chi{\rm HLS}}$ put on FCC crystal lattice.  The skyrmions in the system undergo interactions mediated by the DoFs in the way described, e.g.,  in \cite{PV,HLMR}. What transpires from the calculation focusing on essentials without going into details are (see  \cite{MR-PPNP}):  
\vskip 0.2cm
$\bullet$ {\bf Skyrmion-half-skyrmion ``transition"}
\vskip 0.2cm 

There takes place a topology change from the state of matter with skyrmions to that of half-skyrmions at a density above the nuclear matter density. The transition density $n_{1/2}$ we equate to $n_{BQC}$ is not predicted by the theory. It turns out from the detailed analysis of the data available by astrophysical observation~\cite{MR-PPNP} that
\be
 2\lsim n_{1/2}/n_0 < 4.\label{upper}
  \ee 
In the discussion that follows, we will take this range.   In our approach, the transition density $n_{1/2}/n_0 = 4$ seems to be ruled out~\cite{PREX-II}. It is plausible that further development in astrophysical observations, e.g., the maximum mass of compact stars, might increase the upper limit of $n_{1/2}$ in (\ref{upper}).
  
A characteristic feature of this transition is the resemblance to the pseudo-gap phenomenon in superconductivity.\footnote{Possible pseudo-gap phase was also discussed at high temperature~\cite{zarembo}.} The quark condensate $\Sigma\equiv \la\bar{q}q\ra$, identified as the order parameter in the absence of baryonic matter background,  goes to zero when space averaged (denoted $\bar{\Sigma}$), whereas the pion decay constant $f_\pi$  remains nonzero. Thus the changeover is not a phase transition in the Landau-Ginzburg-Wilson-type sense. For lack of a better term we will continue to refer to it as ``transition" unless otherwise noted. This feature will play the crucial role in formulating G$n$EFT in the class of field theoretic density-functional approach .
\vskip 0.2cm
$\bullet$ {\bf  Soft-to-hard transition in the equation of state} 
\vskip 0.2 cm

One of the most important observations in the skyrmion-to-half-skyrmion  transition is the cusp at $n_{1/2}$ in the symmetry energy $E_{sym}$~\cite{PREX-II}. The symmetry energy, the coefficient of the term proportional to $\zeta^2=\big((N-P)/(N+P)\big)^2$ in the energy per particle of the system $E(n,\zeta)$,  plays the key role in neutron stars with large excess of neutrons. The $E_{sym}$ decreases as it approaches $n_{1/2}$ from below in density, providing attraction, and then after the cusp at $n_{1/2}$, increases rapidly,  thus giving repulsion. Thus the cusp provides the main --  as it turns out --mechanism for the EoS going from soft-to-hard at $\sim n_{1/2}$. This feature  will be found crucial for the maximum mass of neutron stars as well as certain gravity-wave signals coming from merging neutron stars. It turns out also intricately connected to the onset of the pseudo-conformal sound speed at $n\gsim 3n_0$~\cite{PREX-II}.

An interesting observation to make here is that this cusp structure that appears at the leading $N_c$ order in the skyrmion lattice treatment of the symmetry energy is present as an ``inflection" at about the same density as $n_{1/2}$ in phenomenological energy density functional approaches~\cite{df-inflection}

What's given in the skyrmion-crystal simulation is, roughly speaking, a mean-field effect and correlation-fluctuations above the mean field would largely smoothen the cusp, but  the soft-to-hard effect remains unaffected in the EoS. It has been shown that this topological feature can be translated into the nuclear tensor force in  G$n$EFT,  reproducing precisely the cusp structure~ \cite{PREX-II,HLMR,MR-manifestation}.  Another important  consequence of this cusp structure,  is that in going up in density from $n_{1/2}$,  the HLS gauge coupling constant $g_\rho$  is forced to move toward the ``vector manifestation (VM)" fixed point~\cite{HY:PR} at the density $n_{\rm VM}$ at which the vector meson mass vanishes~\cite{suzuki}
\be
m_\rho\sim g_\rho\to 0\ {\rm as}\ n\to n_{\rm VM}\gsim 25 n_0.
\ee
\vskip 0.2cm
$\bullet$ {\bf  Parity-doubling}
\vskip 0.2cm

 The skyrmion-1/2-skyrmion topology change exposes the emergence of another hidden symmetry in strong interactions, namely, the parity doubling. At high density $\gsim n_{1/2}$, the effective nucleon mass deduced from the skyrmion mass tends to converge to  the chiral-invariant mass $m_0\sim (0.6-0.9)m_N$
\be
m_N^\ast\to f_\chi^\ast \sim m_0\not\to  0 \ {\rm as}\  \bar{\Sigma}  \to 0.\label{PD}
\ee
It will be seen below that this implies that the trace of the energy-momentum tensor (TEMT)  becomes  -- in the chiral limit -- a function {\it solely} of $f_\chi\sim m_0$ {\it independent of density} at some density $> n_{1/2}$.  This symmetry is not {\it explicit} in QCD, so one could say it is also emergent. This turns out to have a crucial impact on the sound velocity of the massive stars.

\subsection{Quasi-Free Composite Fermions}\label{quasifree}

\begin{figure}[h]
\begin{center}
\includegraphics[width=4.2cm]{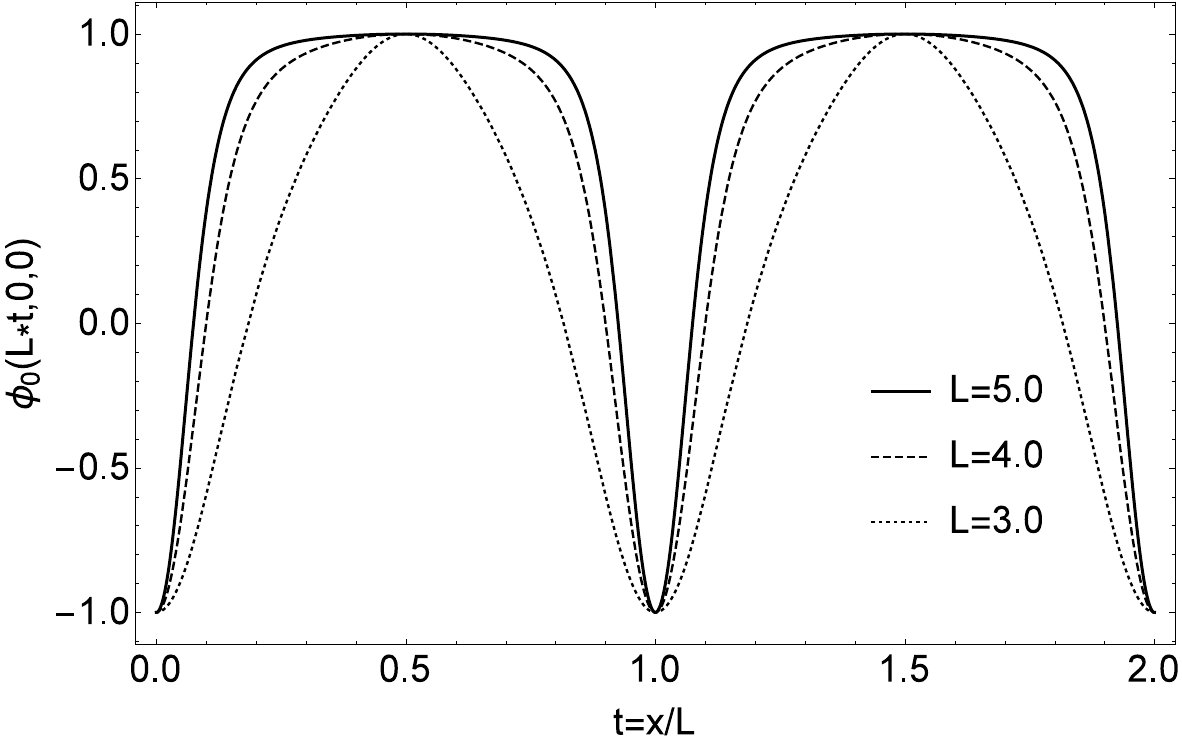}\includegraphics[width=4.2cm]{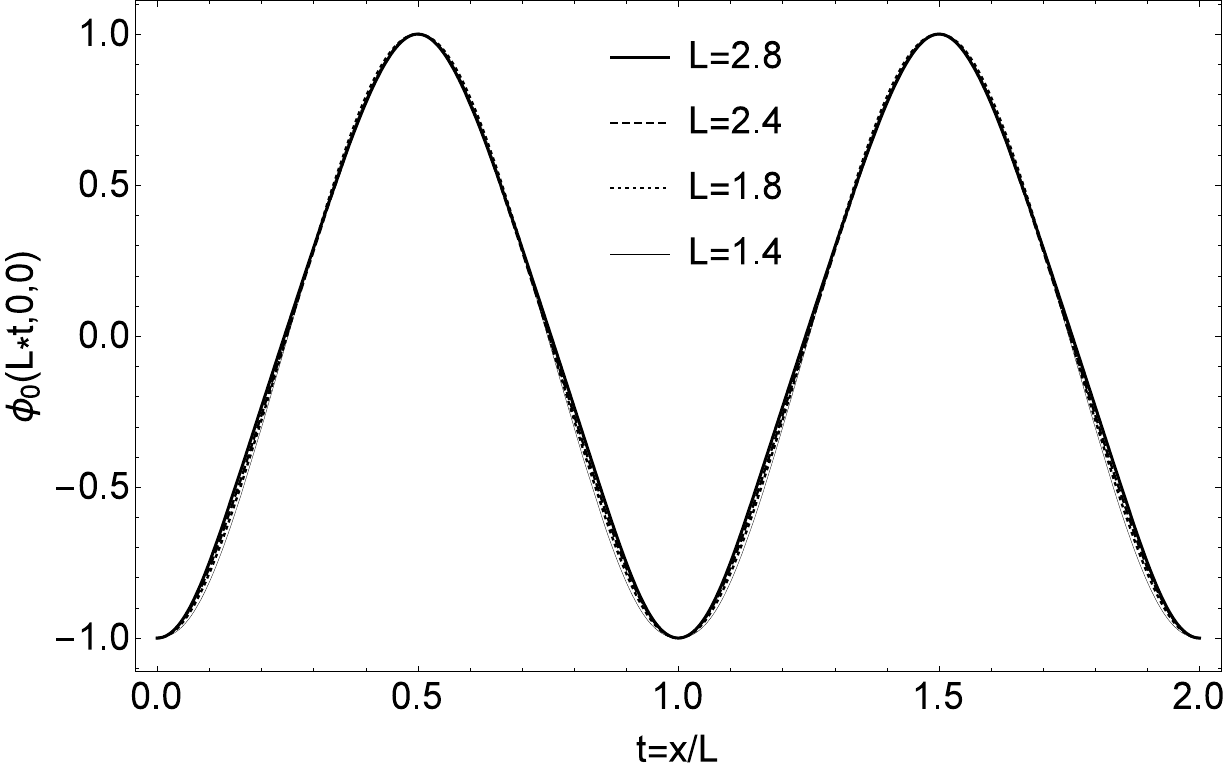}
\includegraphics[width=4.2cm]{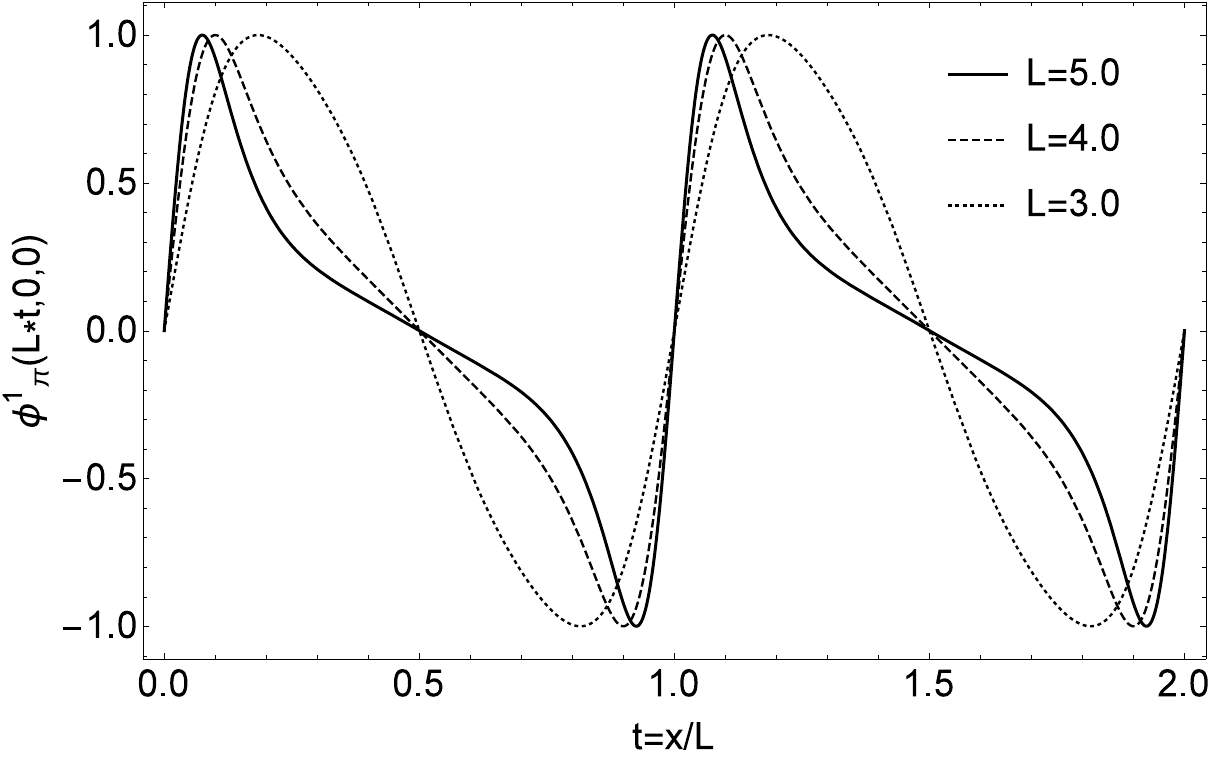}\includegraphics[width=4.2cm]{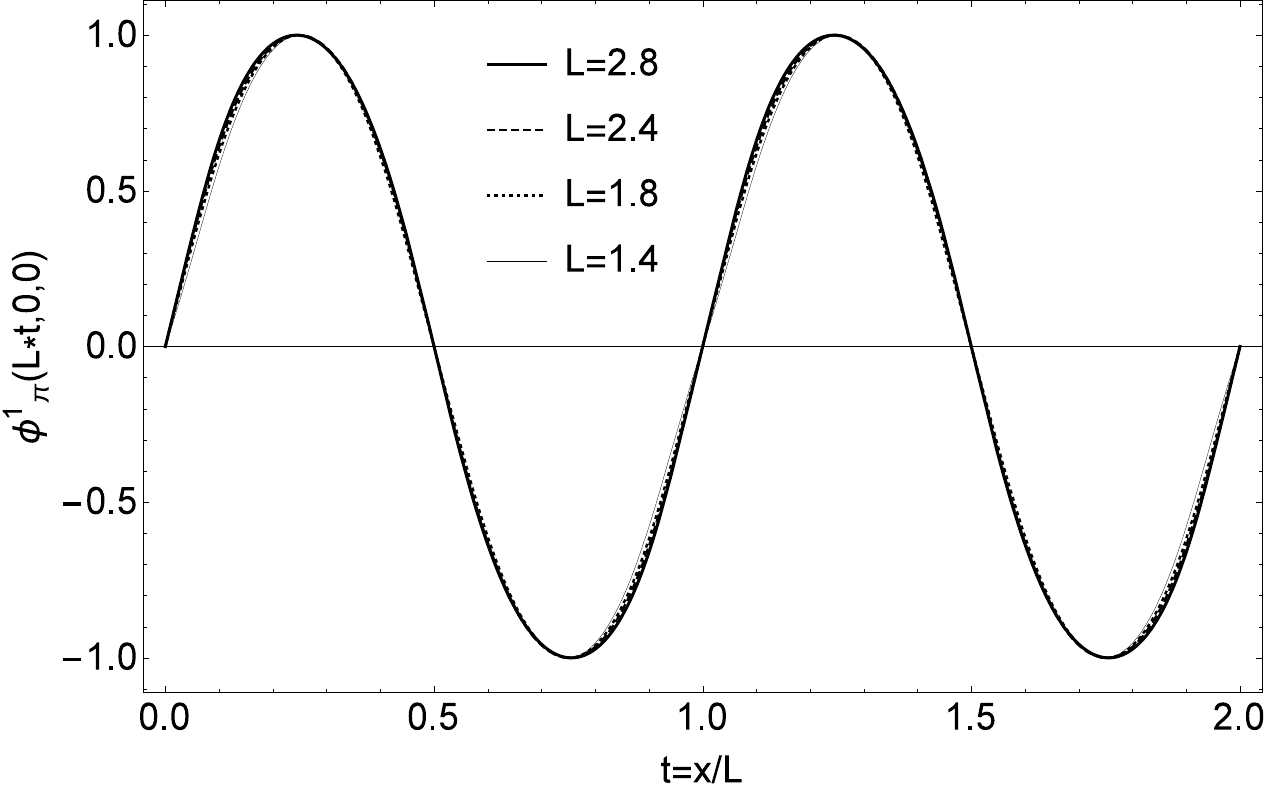}
\caption{  The field configurations $\phi_0$ and $\phi^1_\pi$  as a function of $t = x/L$ along the y = z = 0 line. The left panels correspond to  $n < n_{1/2}$ and the right panels to $n\gsim n_{1/2}$. The half-skyrmion phase sets in when $L=L_{1/2} \lsim 2.9\,{\rm fm}$ where $L$ is the crystal lattice size inversely related to the baryon density of the matter. What is to be noticed is the density independence  of the configurations $\phi_0$ and $\phi^1_\pi$ which engenders what signals ``scale invariance" in the half-skyrmion sector.
 }\label{quasiskyrmion}
 \end{center}
\end{figure}

Though  unproven yet,  it seems very likely that the parity-doubling structure described above is closely related to  that the ``quasi-fermion"\footnote{As will become clear, the fermionic object concerned later can be classed neither as a pure baryon nor as a pure quark. In the absence of a better name, let us just, for simplicity, call it  ``quasi-fermion."}  in the half-skyrmion phase is a nearly non-interacting quasiparticle of baryon number 1.  The first indication came in the Atiyah-Manton approach to skyrmions on crystal lattice~\cite{atiyah-manton}. This turns out to be a highly pertinent observation in the current development in the context described below.  This observation has been discussed in the reviews cited above,  but it is worth recounting it here in terms of insights which were not recognized before. We revisit the result obtained  in Fig.~11 of \cite{PKLMR} reproduced here in Fig.~\ref{quasiskyrmion}. 

What's given here are the field configurations as a function of $t = x/L$ along the y = z = 0 line. They figure in the energy density of the skyrmion matter given by the scale-symmetric HLS Lagrangian ${\cal L}_{\chi {\rm HLS}}$.  We focus on the  chiral field as studied in \cite{atiyah-manton}
\be
U(\vec{x})=\phi_0 (x,y,z) +i\phi_\pi^j (x,y,z)\tau^j.
\ee   
The $\chi {\rm HLS}$ fields lead to similar results.
The field configurations $\phi_{0,\pi}$ correspond  effectively to the the mean fields in $Gn$EFT which transcribed into Fermi-liquid theory~\cite{PKLMR},   can be taken equivalent  to the Fermi-liquid fixed point (FLFP) quantity in the baryonic $\chi$-HLS Lagrangian ${\cal L}_{\psi\chi {\rm HLS}}$ with $\psi$ standing for baryonic field. Those configurations,  varying strongly due to  nuclear correlations with increasing densities in the skyrmion phase at $n < n_{1/2}$ become {\it density-independent in the half-skyrmion phase at $n\gsim n_{1/2}$}. {This means that those configurations representing non-interacting quasiparticles behave scale-invariantly.} This was reflected in the linear density dependence in the cusp for the symmetry energy in the half-skyrmion phase both in the crystal lattice and in $V_{lowk}$-RG calculations which go beyond the FLFP approximation described in \cite{PKLMR} and below.

Particularly interesting is the density-independent configuration $\phi_0$. Since this quantity is proportional to  the pion decay constant $f_\pi$ -- and the dilaton condensate  $f_\chi$ gets locked to   $f_\pi$ going toward the IR fixed point in Crewther's  ``genuine dilaton" scenario~\cite{GDS}, this behavior of $\phi_0$ impacts two important quantities at high density,  first the sound velocity $v_s$ of the compact-star matter,  elaborated below,  and 
second, what is significant at this point is its link to the possible Kohn-Sham energy density functional approach to topology.

One way to understand what we have is to notice that the half-skyrmion is attached with a monopole associated with a hidden $U(1)$ gauge field (say, the $\omega$ field in HLS) whose energy diverges when separated from the other half-skyrmion, but the divergence gets cancelled when the two half-skyrmions are bound, or more precisely, confined~\cite{cho}. The resulting ``composite" skyrmion (baryon) made up of two half-skyrmions in the crystal -- with the kinetic energy suppressed  -- resembles the composite fermion (electron) in the FQHE with the  kinetic energy absent in the limit of large magnetic field~\cite{jain}. What the skyrmion crystal shows  verifies that the quasi-fermions  behave scale-invariantly when the lattice size is varied in the half-skyrmion regime $n\gsim n_{1/2}$.

It should be stressed at this point the quasi-fermion cannot be at or near the IR fixed point. The quasi-fermion that we have is identified as a quasiparticle in Fermi liquid away from the IR fixed point. It is generally considered likely that the Fermi-liquid structure breaks down in electronic systems in the ``unitary limit" at which conformal invariance is present (see \cite{nonfermi} for a recent discussion). This would be the case when the dilaton mass goes to zero in dense-matter systems. When we discuss pseudo-conformal sound speed in compact stars, $v_{pcs}^2\approx 1/3$, we are not dealing with the what's known as ``conformal sound speed" $v^2_{\rm conformal}=1/3$ expected to set in at asymptotic density. The density involved in the stars is far from the IR fixed point density $n_{\rm sIR}$ with non-zero dilaton mass.

Pushing further the analogy to the FQHE, one may wonder whether the anyonic structure encountered in the FQHE as discussed in \cite{jain} has any relevance in the present problem. Indeed there are observations in skyrmion physics obtained with powerful mathematical techniques that there can be $1/q$ (baryon-)charged  objets with $q$ odd integer~\cite{canfora}, and even other more exotic varieties. These may appear to be mathematical oddities, but in our approach they are physical.  We will return to this issue in the second part of this note where fractional quantum Hall droplets (pancakes or pitas)  could figure at high density.

\section{Translating topological inputs into effective field theory G$n$EFT}
%
%

Given the topological inputs extracted from the skyrmion-half-skyrmion transition, the next step is to incorporate them into the EFT.  To build the EFT concerned as an ``analog" to the KS-DFT in the FQHE in \cite{jain} in the strategy to map topological properties to an EFT, we introduce baryon fields explicitly and couple them to $\chi$HLS fields chiral-scale symmetrically to ${\cal L}_{\chi{\rm HLS}}$. Let us denote it  ${\cal L}_{{\psi\chi}{\rm HLS}}$ with $\psi$ standing for baryons.  It is of the form
\begin{eqnarray}
\mathcal{L}_{{\cal L}_{{\psi\chi}{\rm HLS}}}
&=& \bar{N}i\gamma^{\mu}D_{\mu}N - hf_{\pi}\frac{\chi}{f_{\chi}}\bar{N}N + g_{v\rho} \bar{N}\gamma^{\mu}\hat{\alpha}_{\parallel \mu}N\nonumber\\
&+ & g_{v0} \bar{N}\gamma^{\mu}\mbox{Tr}\left[\hat{\alpha}_{\parallel \mu} \right]N
{}+ g_{A}\bar{N}  \gamma^{\mu}\hat{\alpha}_{\perp \mu}\gamma_{5} N .
\label{BHLS}
\end{eqnarray}

In the presence of explicit baryons, both the topological and ``homogeneous" Wess-Zumino terms~\cite{HY:PR}\footnote{Going toward the  $\eta^\prime$ ring structure as the density is increased, the homogeneous Wess-Zumino  (hWZ) term  becomes ``hidden WZ" term of \cite{karasik2} with the coefficients vector-dominated and the FQH droplet ``visible."} in the mesonic Lagrangian ${\cal L}_{{\chi}{\rm HLS}}$ from which the skyrmions are built are absent. This is of course familiar in $s$ChEFT with the standard (nuclear) chiral Lagrangian with or without strangeness. 
\subsection{Density Functional via Fermi-liquid Fixed-Point Theory}
We confine to one unique Lagrangian defined {\it only} with the {\it relevant hadronic} variables ${\cal L}_{{\psi\chi}{\rm HLS}}$ and  eschew ``hybridization" with non-hadronic degrees of freedom.  The topology change will be encoded in the parameters of the Lagrangian ${\cal L}_{{\psi\chi}{\rm HLS}}$ that change at the density $n_{1/2}$.   Below the transition density, the Lagrangian endowed with the well-defined scaling of the pion and dilaton decay constants as dictated by the matching with QCD -- i.e., via correlators --  as proposed a long time ago~\cite{BR91} should reproduce $s$ChEFT.  How well the predicted results fare with the established data of nuclear matter  has been extensively reviewed (e.g., \cite{MR-PPNP}).    Of course with the extreme simplification, one cannot hope to match the precision enjoyed by $s$ChEFT treated at high chiral orders.  However the property of the equation of state (EoS), in particular the symmetry energy $E_{sym}$ approaching $n_{1/2}$ from below, does have certain potentially nontrivial features associated with the properties such as tidal polarizability (TP) measured in the recent gravity waves which differs from the prediction of $s$ChEFT.  

As expected for possible hadron-quark continuity,  the parameters of ${\cal L}_{{\psi\chi}{\rm HLS}}$ are drastically affected by the topology change at $n_{1/2}$. They can differ qualitatively in the half-skyrmion phase from what was predicted in \cite{BR91} and hence from naive extrapolation to $n\gsim n_{1/2}$ in $s$ChEFT. The most important impact  is on the property of the bound half-skyrmions behaving as a scale-invariant quasi-fermion described in Sec. \ref{quasifree}.  It is represented in the Lagrangian ${\cal L}_{{\psi\chi}{\rm HLS}}$ as an effective baryonic field $\psi$ with its physical properties -- the mass and coupling constants -- dictated by the topological properties of the half-skyrmion phase, i.e., the near density independence of the effective mass and coupling constants and suppressed kinetic energy.  Equally important is the (assumed) composite (HLS) gauge symmetry \`a la Suzuki theorem with the VM fixed point with the vanishing vector mass at $n_{VM}\gsim 25 n_0$. It makes the vector field coupling to the quasi-fermion strongly weakened, leading to what we propose as the emergence of (pseudo-)conformal symmetry.

It should in principle be feasible to develop chiral-scale symmetric EFT in a parallel to high-order $s$ChEFT successful in nuclear physics at density $\sim n_0$. The cutoff could be set at $\Lambda_{Gn{\rm EFT}}\gsim m_V$.  This feasibility in HLS was already discussed in \cite{HY:PR} and initiated in $\chi$HLS~\cite{LMR}.   Unfortunately it is not in a form to perform high enough order scale-chiral perturbation calculations. Here we will resort to a strategy resembling what is done in the FQHE~\cite{jain}, a sort of an application to dense nuclear matter of  density functional theory consistent with Hohenberg-Kohn  theorem~\cite{Hohenberg-Kohn}. 

In a nut-shell, the chain of reasonings goes as follows.

As well recognized in nuclear theory circles, the relativistic mean field theory as first formulated in Walecka's linear model~\cite{walecka} belongs to the class of density functional approaches. It has been extensively exploited in terms of the Kohn-Sham density functional  in nuclear structure studies. Furthermore in conjunction with $s$ChEFT, {\it ab initio} calculations in Kohn-Sham density functional are being explored, with the possibility of doing precision nuclear structure calculations.  All these efforts are however limited at present to the density regime $\lsim n_{1/2}$.

It is also known, though perhaps not so widely,  that the Walecka model captures Landau Fermi-liquid theory~\cite{matsui}. Next a chiral Lagrangian implemented with the HLS mesons is established~\cite{friman-rho,FR-song} to lead to the Wilsonian renormalization-group approach to Fermi-liquid fixed-point theory~\cite{shankar,polchinski}. It follows then that the Lagrangian  ${\cal L}_{{\psi\chi}{\rm HLS}}$ with the parameters encoding the topology change and matched to QCD in medium  is expected to give in the mean field a highly powerful Fermi-liquid fixed point theory that can access densities $\gsim n_{1/2}$. The reliability of the large $\bar{N}$ approximation (where $\bar{N}=k_F/(\Lambda_{FS}-k_F)$ with $\Lambda_{FS}$ the cutoff on top of the Fermi surface)  the Fermi-liquid fixed point approximation indicates the validity of the mean field for densities near and higher than $n_0$ and more specially for high density~\cite{walecka}. In fact it  is feasible to go beyond the mean-field by including $1/\bar{N}$ corrections taken into account in what is known as $V_{lowk}$RG. 
\subsection{``Quenched $g_A$" as Precursor to Emergent Scale Symmetry at $n\gsim n_{1/2}$}
In nuclear physics, the $V_{lowk}$  renormalization-group method has been applied to go  beyond the FLFP approximation~\cite{PKLMR} along the line of the Wilsonian renormalization-group strategy~\cite{polchinski,shankar}. What happens as the IR fixed point density $n_{\rm sIR}$  is approached has not been addressed. This density is never likely arrived at in compact stars where the dilaton mass is not zero, but it will become relevant at higher densities considered in what follows. In a recent development in condensed matter physics, the approach to soft modes -- gapless phase -- on top of the Fermi surface going beyond what amounts to the  FLFP theory is being formulated by nonlinear bosonization of the quasiparticles on the Fermi surface leading to an order-by-order counting of the beyond-the-FLFP higher-order terms~\cite{coadjoint-orbit}. This would provide a systematic calculation of higher-order terms resembling that of chiral perturbation theory leading to sChEFT, illuminating the role of heavy DoFs in G$n$EFT.  An approach of this type may give possible corrections to the solution to the quenched $g_A$ problem as well as the pseudo-conformal property of massive stars, both of which are discussed below.

The ultimate goal of G$n$EFT is to access the high density regime relevant to massive compact stars. Most intriguingly there is a hint however already at low density $n\lsim n_0$  to what might be happening with the emergent scale symmetry at high density. Without going into details here (readily found in the reviews, such as \cite{MR-PPNP}), we illustrate a case at low density $\sim n_0$ which shows a predictive power not shared (up-to-date) by $s$ChEFT, with a close link to what takes place at $n\gsim n_{1/2}$. It is found to provide a simple and  elegant resolution to a long-standing mystery lasting several decades of the ``quenched" $g_A$ observed in nuclear Gamow-Teller transitions in light nuclei~\cite{MRgA}.  The key element of the solution is that at the mean field level, i.e., at the Fermi-liquid fixed point, the superallowed Gamow-Teller transition described in G$n$EFT is precisely given by the soft-pion and soft-dilaton theorems, namely, the Goldberger-Treiman relation involving the density-dependent Landau Fermi-liquid fixed-point  parameters $F_1$ that enter in the Landau effective mass. 

The quantity identified as the Fermi-liquid fixed point axial constant $g_A^{\rm Landau}$ in EFT approach is found to encode ``hidden scale invariance" emerging from strong nuclear correlations~\cite{multifarious}. As discussed in \cite{MRgA} it can be equated to the effective $g_A$ given by the ``extreme single-particle shell-model (ESPSM)"  defined for doubly closed-shell nuclei corresponding to the ``scale-symmetric effective $g_A$" (denoted as $g_A^{\rm ss}$). To render a precise physical meaning to this quantity, imagine that one were able to obtain the -- up-to-date non-existent -- {\it exact} wave functions   of the parent and daughter states involved in the superallowed (zero momentum transfer) Gamow-Teller transition. Now the exact transition matrix element in the doubly closed-shell nuclei would then be given by the Gamow-Teller matrix element given by the ESPSM multiplied by the constant $g_A^{\rm ss}=1$ modulo scale-anomaly correction factor $q_{ssb}$~\cite{multifarious}. This means that the total nuclear correlations are encapsulated in the constant $g_A^{\rm ss}$ multiplying the ESPSM matrix element $M^{\rm ESPSM}/g_A$ where $g_A$ is the axial constant in neutron decay.  In terms of Landau Fermi-liquid fixed point effective field theory, it corresponds to the quasiparticle sitting on top of the Fermi surface making the superallowed Gamow-Teller transition. Therefore $g_A^{\rm Landau}=g_A^{\rm ss}=1$. Now the expression for the Fermi-liquid fixed point $g_A^{\rm Landau}$ which corresponds to the Ward identity is essentially the nuclear matter version of the Goldberger-Treiman relation established to hold within a few \% accuracy in matter-free space~\cite{FR-song}.\footnote{We remind what was established a long time ago.  The expression for $g_A^{\rm Landau}$ is the axial current counterpart of the Ward identities for the EW currents. The EM components are the nuclear orbital gyromagnetic ratios which give the iso-scalar $g_l^{(0)}=1$ agreeing perfectly with the Kohn theorem and the isovector $g_l^\tau$ agreeing with the Migdal formula given in terms of the Landau-Migdal Fermi-liquid quasiparticle interaction parameters. These illustrate the power of the EFT approach to Landau Fermi-liquid theory in nuclear matter.} What this means is that  the more correlations were taken into account, the closer the axial constant that multiplies the single-particle Gamow-Teller operator would come to the {\it unquenched} value $g_A=1.27$. In very light nuclei where powerful numerical techniques are available, this is in fact what's found with possible many-body (exchange-current) contributions amounting to $\lsim$ 2\%~\cite{Quantum Monte Carlo}. 

In short, that $g_A^{\rm ss}\approx 1$ in nature would indicate an emergence of quantum scale invariance as suggested in \cite{multifarious}. That the putative dilaton mass in nuclei is $\sim 600$ MeV, so scale symmetry is certainly broken -- at least spontaneously, so the quantum scale invariance must be hidden at low density.  Deviation of the factor $q_{\rm ssb}$ from 1 would indicate that quantum scale invariance is broken in nature.  A more precision re-measurement of the RIKEN experiment of the superallowed GT transition in the doubly magic nuclei discussed in \cite{MRgA}  would be valuable on this issue. 

At the dilaton-limit fixed point at high density that we discuss below, however,  $g_A$ does approach 1.  What is particularly interesting is that  the quantum scale invariance hidden at low density, presumably responsible for the  quenching of $g_A$,  seems to emerge in the ``pseudo-conformal" sound velocity $v_s^2/c^2\approx 1/3$ in massive compact stars as discussed below. Whether this result holds beyond the FLFP approximation could be addressed in this nonlinear bosonization technique~\cite{coadjoint-orbit}. One could also address neutrinoless double $\beta$ decay where the momentum transfer could involve $\sim 100$ MeV.

\section{Going toward massive compact-star matter}
As long as the density of the core of massive stars is not in the vicinity of the IR fixed point\footnote{At the IR fixed point, due to the soft scalar mode, the Fermi-liquid structure must break down.}, we assume  that the notion of the Fermi-liquid fixed point applies to the density regime $n\gsim n_{1/2}$ as it does at $n\lsim  n_{1/2}$.  One can make two simple calculations of the EoS to go toward the density relevant to massive compact stars. Given the paucity of the trustful knowledge about the structure of the state involved, the guidance available  is the presumed constraints provided by the symmetries assumed to underlie the dynamics: The HLS (of the composite gauge symmetry~\cite{suzuki}) with the vector manifestation (VM) density $n_{\rm VM}\gsim 25 n_0$ and the HSS with the IR fixed point at $n_{\rm sIR}$ which we assume is near $n_{\rm VM}.$\footnote{In the present framework there is no compelling reason to believe that $n_{\rm VM}\simeq n_{\rm sIR}$. The only thing one can say -- and assumed here -- is that both $n_{\rm (VM,sIR)}$ are higher than what's relevant to the maximum density supported by massive compact stars stable against gravitational collapse.} 
\subsection{Dilaton-Limit Fixed Point}\label{DLFP}
Consider  ${\cal L}_{{\psi\chi}{\rm HLS}}$ for $n\geq n_{1/2}$ with the parameters of the Lagrangian taken density-dependent, but totally arbitrary, {\it unconstrained} by the topology change discussed above. Assume that the mean-field approximation holds at $n\gsim n_{1/2}$ in the sense defined in the large $N_c$ and large $\bar{N}\sim k_F$ limit. Now what we would like to do is to take what corresponds to going toward the IR fixed point of the GDS.  This can be done by following  Beane and van Kolck~\cite{bira}: First do the field re-parametrization  ${\cal Z}=U\chi f_\pi/f_\chi=s+i\vec{\tau}\cdot\vec{\pi}$ in ${\cal L}_{{\psi\chi}{\rm HLS}}$, have the Lagrangian treated in the mean field  and take the limit ${\rm Tr} ({\cal Z}{\cal Z}^\dagger)\to 0$. This limit is referred to as ``dilaton-limit fixed point" (acronymed DLFP). Two qualitatively different terms appear from this manipulation:  one is regular and the other singular in the limit. The singular part is of the form
\be
{\cal L}_{\rm sing} &=& (1- g_A){\cal A} (1/{\rm Tr} \big({\cal Z}{\cal Z}^\dagger)\big)\nonumber\\ 
&+& (f_\pi^2/f_\chi^2 -1) {\cal B} \big(1/{\rm Tr} ({\cal Z}{\cal Z}^\dagger)\big).
\ee
The first (second) term is with (without) the nucleons involved. The requirement that there be no singularities leads to the ``dilaton-limit fixed point (DLFP)" constraints
\be
g_A\to g_A^{DL}=1\label{gAto1}
\ee 
and 
\be 
f_\pi\to f_\chi \neq 0.\label{fpifchi}
\ee
We have denoted the $g_A$ arrived at the DLFP as $g_A^{\rm DL}$ to be distinguished from the $g_A^{\rm Landow}$ in \cite{MRgA} arrived at the Landau Fermi-liquid fixed point at $n\sim n_0$. It turns out that the $\rho$ meson decouples from fermions in dense matter going toward the DLFP even though the gauge coupling $g_\rho\not\to 0$~\cite{PKLMR}.  Therefore the $\rho$ meson drops out before reaching the VM fixed point.

These ``constraints" are the same as what are in the genuine dilaton properties approaching the IR fixed point~\cite{GDS}. This suggests that  the topological characteristics of the half-skyrmion phase are consistent with the GDS.  We should mention that there are other constraints in dense matter associated with  the DLFP, among which highly relevant to the EoS at high density is the ``emergence" of parity doubling in the nucleon structure mentioned above.   
\subsection{Emerging Pseudo-Conformal Symmetry}\label{EPCS}
Next consider ${\cal L}_{{\psi\chi}{\rm HLS}}$, in contrast to what's done above, with its parameters {\it constrained} by the skyrmion-half-skyrmion topology change at $n_{1/2}$. As shown, in the half-skyrmion phase,  as the system flows toward the IR fixed point (or  perhaps equivalently the VM fixed point) -- although $n_{1/2}\ll n_{\rm sIR}$, the parameters set in {\it precociously} as
\be
f_\pi\to f_\chi,
\ee 
and
\be
\ m_N\to f_\chi\propto \la\chi\ra \to m_0\label{mN*}
\ee
leading to the parity doubling (\ref{PD}).

In the mean field, that is,  in the Landau Fermi-liquid fixed-point approximation in G$n$EFT, the energy-momentum tensor is easily calculable. It comes out to be~\cite{PKLMR}
\be
\la\theta^\mu_\mu\ra=4V(\la\chi\ra) -\la\chi\ra\frac{\del V(\chi)}{\del\chi}|_{\chi=\la\chi\ra}\neq 0
\ee
where all the conformal anomaly effects (and also quark mass terms) are lumped into the dilaton potential $V(\chi)$. 
Thus  $\la\theta_\mu^\mu\ra$ is a function of {\it only} $f_\chi$ which does not depend on density (far below $n_{\rm VM}$) via (\ref{mN*}). It follows that
 \be
 \frac{\del}{\del n} \la\theta_\mu^\mu\ra=0
 \ee
and hence
\be
 \frac{\partial \epsilon(n)}{\partial n}(1-3v_s^2/c^2)=0
\ee
where $v_s^2=\frac{\del P(n)}{\del n} (\frac{\del\epsilon}{\del n})^{-1}$ is the sound velocity and $\epsilon$ and $P$ are respectively the energy density and the pressure. It is fair to  assume that there is no Lee-Wick-type anomalous nuclear state at the  density involved, so $\frac{\partial \epsilon(n)}{\partial n}\neq 0$. Therefore we have
\be
v_{pc:s}^2/c^2\approx 1/3.\label{pcs}
\ee 
Note that this is not to be identified with the ``conformal sound velocity" $v_s^2/c^2=1/3$ expected at asymptotic density.  The trace of the energy-momentum tensor is not zero at the compact-star density, so we call this pseudo-conformal (PC) sound velocity $v_{pc:s}$. 

We should mention at this point a surprising observation  in the $V_{lowk}$RG calculation that takes into account higher $1/\bar{N}$ corrections  in $E/A$  of the $A$-nucleon ground state. The emergence of the PC symmetry is found to be intricately tied to whether the VM fixed point is lodged in the vicinity of the core of massive stars or at a much higher density. If it is taken far above the core density, say, at $n_{\rm VM}\gsim 25 n_0$, then $E/A$ at $n\gsim n_{1/2}$  can be very accurately reproduced by the two-parameter formula~\cite{PKLMR,MR-PPNP}
\be
E/A= -m_N + X^\zeta (n/n_0)^{1/3}+Y^\zeta (n/n_0)^{-1}\label{E/A}
\ee 
where $\zeta=(N-Z)/(N+Z)$ and $X$ and $Y$ are the constants to be fixed by equating (\ref{E/A}) to the $E/A$ given in $V_{lowk}$RG at $n=n_{1/2}$ by continuity in the chemical potential and pressure.  $X$ and $Y$ depend on where $n_{1/2}$ is located. One can show that $\la\theta_\mu^\mu\ra$ is a constant independent of density for any values of $X$ and $Y$. This then gives rise to the PC sound velocity (\ref{pcs}). On the contrary if if $n_{\rm VM}$ were taken at $n\sim 6n_0$, say, in the center of massive stars, then the sound speed $v_s^2/c^2$ could not set in at the PC value in the range of star density but would exceed 1/3. This brings another surprise: That {\it  the PC symmetry is intimately tied to the VM property of hidden local symmetry.} Why this is so remains un-understood.

Note that (\ref{pcs}) is an approximate equality (with non-vanishing TEMT), not the equality which would hold at the asymptotic density $\gg n_{1/2}$.  In the density regime concerned, $n\lsim 7 n_0$, there can however be deviations due to the quark mass term,  and also higher-order terms of the anomaly-induced symmetry breaking involving the anomalous dimension $\beta^\prime$ that could make the sound speed deviate from (\ref{pcs}). However there are no reasons to suspect that the corrections would make $v_s$ deviate appreciably from $v_{pc:s}$.

What is most glaringly different between the prediction of the G$n$EFT and that of {\it all} other models in the literature is the onset of the pseudo-conformal (PC) sound speed (\ref{pcs}) at a relatively low density $\sim 3n_0$ which stays more or less constant up to the central density $\sim 6n_0$ of massive star $M_{max}\lsim 2.3 M_\odot$.  

As already stated, as far  as we are aware, there are no observables so far measured with which the results of the G$n$EFT (including the recent GW observables) are at odds~\cite{YLM}. Because of the change of parameters of the Lagrangian  ${\cal L}_{{\psi\chi}{\rm HLS}}$ controlling the EoS obtained from G$n$EFT, the most drastic of which is the cusp in $E_{sym}$ at the leading order, there occur strong fluctuations in the density regime $\sim (2-4)n_0$ at which the topology change takes place. This gives rise to a spike in the sound velocity in that region  after which the sound velocity $v_{pcs}^2/c^2$ stabilizes  quickly to 1/3 above $\sim 3n_0$. The strength of the spike below  the transition region can vary depending on the value of $n_{1/2}$. It can even overshoot the causality limit, for instance,  for $n_{1/2} \gsim 4 n_0$\footnote{This was the result that set the upper limit of $n_{1/2}$ to 4 $n_0$ in the bound (\ref{upper}).}. This strong enhancement in the sound speed going over the normal hadronic-to-non-hadronic crossing can also be seen with the transition mediated by hadronic-quarkyonic continuity~\cite{quarkyonic}. Thus this aspect of the sound speed could very well depend on how the changeover from hadronic to other forms of the state of matter takes place.  This of course would be too difficult an issue to accurately sort out in the (over)simplified description. What is less unambiguous is the precocious onset of the PC sound velocity.  

The robust takeaway from this result is that in the way the PC symmetry permeates from low density ($\lsim n_0$) in the $g_A^{\rm Landau}\approx 1$ to high density ($ > n_{1/2}$) in the $g_A^{\rm DL}=1$,  the PC sound velocity simply reflects the precocious emergence of the same PC symmetry. Among others it predicts that in the core of massive stars at a density $\sim 6 n_0$,  the objects found there are the composite quasi-fermions of bound half-skyrmions.  

The question then is: Are these quasi-fermions unrelated to what might be described as ``deconfined quarks"?  

It has recently been argued in \cite{deconfined}, based on detailed analyses combining astrophysical observations and  theoretical calculations, that the matter in the core of maximally massive stars exhibits the characteristics of  ``deconfined phase" and suggests that the fermions residing in the core are most likely ``deconfined quarks."  The prediction of G$n$EFT differs from this interpretation: The objects found in the core are neither purely quarks nor purely baryons but quasi-fermions of the confined half-skyrmions~\cite{core}. The resemblance is however uncanny if one compares the predictions $P/\epsilon$ where $P$ is the pressure and $\epsilon$  is the energy-density as function of density $n$ and the polytropic index $\gamma=d({\rm ln}P)/d({\rm ln}\epsilon)$ made in the description given above with the analysis of \cite{deconfined}. This, we suggest, is the reflection of the topology change, a.k.a., baryon-quark continuity.  We now turn to this issue in terms of what might be called ``hadron-quark duality."
\section{Hadron-Quark Duality\\ And Cheshire Cat Phenomenon}
The description given above involves composite fermions made of half-skyrmions in some sense ``masquerading" as fractionally charged quarks.    As noted above, however, half skyrmions are not the only objects that skyrmions can turn into. There could be other fractional objects such as mentioned  in \cite{canfora} and others to be mentioned below.

Thus far we have ignored the possible role that the FQH droplet might play in dense matter. While the skyrmion description applies to the octet baryons coming from the octet mesons for the flavor $SU(3)$,  there is no skyrmion for the $U(1)$ meson $\eta^\prime$, i.e., the ``dichotomy problem." A possible solution to this dichotomy is that at large $N_c$ limit, the baryon coming from the $\eta^\prime$ is a FQH droplet or more appropriately ``pancake" described  in Chern-Simons field theory~\cite{komargodski}, not a skyrmion. In nature, the $\eta^\prime$ associated with the chiral anomaly is massive given that $N_f\ll \infty$, so could be ignored in low-energy/density dynamics of baryons. Indeed there seems to be no  indication that it figures directly at least at low density\footnote{It does however figure, though indirectly, in the proton's suppressed flavor-singlet axial coupling-constant $g_A^{(0)}<< 1$~\cite{FSAC} explained as due to the color anomaly~\cite{colorleakage}  we will return to below.}. However the question arises as to what happens at high density (and also at high temperature) where ``deconfined" quarks are naively expected  to intervene. There is a highly original and provocative  -- and,  viewed from the point of view of our approach, compelling -- argument~\cite{karasik1,karasik2,kitano} that indeed the FQH pancake is {\it essential} at some high density (and/or temperature) in the vicinity of chiral restoration with the possible restoration of $U_A(1)$ symmetry linked with the dropping $\eta^\prime $ mass~\cite{UA1sym}. It has, thus far unexplored, implications on how the vector mesons $\rho$ and $\omega$ in $\chi$HLS behave near the chiral transition~\cite{kitano}, for instance in  heavy ion physics, with a possible paradigm change in the field. This could also be relevant to the inescapable question in confronting the theory with experiments in massive stars, say, as ``deconfined quarks"~\cite{deconfined} or pseudo-quarkonic phase suggested in the literature.  

This brings us to the old question of what heavy nuclei are in QCD, an issue hotly discussed in 1970's.

Consider $N_c=3$ ``confined quarks"  in, say, the MIT bag for a nucleon  in nuclear matter of mass number $A$. When squeezed in dense matter, as the bags overlap, one can visualize the quarks ultimately percolating from one bag to another bag and then coalesce into one big bag of $N_c\times A$ quarks. In 1970s, this is the way some nuclear theorists thought of the $^{208}$Pb nucleus as 624 quarks interacting via perturbative QCD  confined within a giant bag. Such a picture was not -- and still is not -- a feasible one for the reason by now well-known at least for low density. Even so, incorporating the MIT bag structure with asymptotic free interactions at high density could make at least qualitative sense at asymptotic densities.  Indeed many papers have been written where low-density hadronic description is hybridized with MIT bag description at increasing density.  They typically involve phase transitions. We cite just a couple of the most recent of them~\cite{MIT} where other relevant references can be found.

One possible alternative was put forward for nuclear dynamics at low energy (and low density) by what was called by ``Cheshire Cat Principle"~\cite{CCP} whereby $N_c$ quarks in a bag transform into a topological soliton, skyrmion, so the quarks disappear into the ``smiles" of the Cheshire Cat with the solitons interacting via fluctuating meson exchanges, in the way Weinberg admitted  as what ``nuclear physicists knew what they were doing" before the advent of the $s$ChiEFT as prescribed in the Folk Theorem.
\subsection{``Infinite Hotel" for $N_f\geq 2$: Skyrmions}
What takes place can be imagined as a quark in a ``jail" trying to escape from the jail, fully occupied,  like the filled Dirac sea\footnote{This ``jail-break" scenario is beautifully described in \cite{jail-break}. Actually $N_c$ quarks are involved but we focus on only one of them.}.  A massless quark swimming on top of the sea, say,  to the right  in one spatial dimension\footnote{The argument can  be straightforwardly extended to 3 spatial dimensions.},  in an attempt to escape the jail,  gets blocked at the ``jail wall," so is unable  to escape.  It cannot swim back on top of the Dirac sea,  because chiral symmetry forbids it. But it can plunge into the Dirac sea which is feasible, because the Dirac sea is infinite, and swim back to the left inside the sea. This infinite Dirac sea can be likened to an  ``infinite hotel (IH)"~\cite{jail-break}.  This exploitation of the infinity  is  a quantum effect known as ``quantum anomaly."

There is one serious problem in this scenario, however. The fermion (baryon) charge carried by the quark disappears into the Dirac sea, so  the baryon number is apparently ``violated" in the process.  In QCD, the baryon charge is absolutely conserved, so the fermion charge cannot disappear. Here takes place a miracle.  The fermion charge is relayed to the ``pion"  that clouds the outside wall,  with the pion (boson) turning into a baryon (fermion). This is by now the well-known story of skyrmions in (3+1)D mathematically characterized by the homotopy group $\pi_3 (S^3)=\mathcal{\cal Z}$ for the $N_f\geq 2$ systems.

This IH phenomenon can be considered to involve two domains, one the quark-gluon one and the other the hadronic one. There are two modes of a global symmetry, i.e., chiral symmetry, involved:  Wigner-Weyl (WW) mode inside the bag and Nambu-Goldstone (NG) mode outside the bag. Therefore the jail wall can be taken as a thin ``domain wall" that delineates two vacua. This is the ``jail-break" scenario for the $N_f=2$ (i.e., proton and neutron) case. 

The upshot is that the leaking baryon charge is taken up by the pion as a soliton. So in nuclear physics, we argue that for the given soliton chiral angle  $\theta(R)$, the leaking baryon charge $1-\theta (R)/\pi$ (in 1 spatial dimension) is lodged in the skyrmion cloud while the rest of the  charge  $\theta (R)/\pi$ remains in the bag, yielding the total baryon number 1 for a single baryon.  When  the bag is infinite the whole baryon charge is lodged  inside the bag, while when the bag shrinks to zero size the whole baryon charge goes into the skyrmion cloud. So the size of the bag has no meaning for the property of the quark. The confinement size idensified with the bag size $R$ is therefore an unphysical quantity. One can think of this process as the pion fields giving rise to the baryons as solitons. This is what is referred to as ``Cheshire Cat Phenomenon" or ``Cheshire Cat Principle (CCP)"~\cite{CCP}. This   is akin to the disappearance of the Cheshire Cat in ``Alice in the Wonderland" with the baryon number playing the role of the cat's smile.  In fact it could be more appropriate to identify this phenomenon as a gauge artifact and formulate a gauge theory for the phenomenon~\cite{CC-gauge}.

This discussion of the CC ``smile" applies straightforwardly to (3+1) dimensions. It has indeed been verified by Goldstone and Jaffe~\cite{GJ}  in terms of the spectral asymmetry $\eta(s)$ (defined in (\ref{SA}) below) which renders the baryon charge lodged inside the bag for a given chiral angle $\theta (R)$. The fractionalization of the baryon charge is exact thanks to the topology involved. In (1+1) D, an exact bosonization allows an in-principle CCP also for non-topological processes.    But in the absence of bosonization, such exact CCP does not exist in nuclear processes  in (3+1)D -- except for the topological quantity, so much of what one can say of  the processes in nature that are not topological is at best approximate. 
\subsection{No Infinite Hotel for $N_f=1$ Baryons}
The IH scenario discussed above famously turns out not to work when the number of flavors is one. This is because $\pi_3 (U(1))=0$. One then wonders whether there is no soliton for baryon coming from the flavor singlet meson $\eta^\prime$. As mentioned, this puzzle was recently resolved by ideas developed in  condensed matter physics by Komargodski~\cite{komargodski} who suggested that the $\eta^\prime$ can turn into a flavor singlet baryon -- denoted from here on as $B^{(0)}$ -- as a fractional quantum Hall (FQH) droplet. At first sight this  FQH droplet (pancake or pita) is unrelated to the usual skyrmion corresponding to the nucleon and hence the dichotomy.

There are two questions raised regarding this dichotomy~\cite{dichotomy} between the skyrmions and the FQH droplets. The first is: Is there any relation between the two topological objects, the FQH droplet for $B^{(0)}$ and the skyrmion for nucleons?  The second is: Is the phenomenon of the FQH droplets relevant to the EoS at high density? Both questions are in some sense related.
\subsubsection{Baryon for $N_f=1$}
Let us first discuss whether Komargodski's FQH pancake model can be given a formulation in terms of a Cheshire Cat phenomenon. This rephrases what was done in \cite{MNRZ}. 

Suppose the quark in the bag is of $N_f=1$ in the jailbreak scenario.  Let the quark be coupled at the wall $x=R$ to the flavor-singlet meson $\eta^\prime$. Again the confinement leads to the breaking of the baryon charge and gives rise to an anomaly, but since $\pi_3 (U(1))=0$, it {\it cannot} go into the infinite hotel because the topology does {\it not allow} it.  We will come later to how and why the topology might ``dictate" the flow. So where does it go?
The answer~\cite{MNRZ} is that the quark moving in the $x$ direction is allowed to escape by flowing in the $y$ direction and go into a  2d quantum Hall-type pancake, taking care of the anomaly generated by the boundary condition\footnote{Below this boundary will be identified with a thin domain wall.} and keeping the baryon charge conserved. This is known as the ``anomaly in-flow" mechanism leading to the Chern-Simons topological term ~\cite{callan-harvey},  which in 3-form reads 
\be
\frac{N_c}{4\pi}\int_{2+1} ada\label{cs}
\ee
where  $a_\mu$  is the Chern-Simons field which is to capture strong correlations in QCD -- and will be identified later with the $U(1)$ field in HLS, namely the $\omega$ meson~\cite{karasik1,karasik2}. The Cheshire Cat Principle, if held, would imply  that the baryon charge leaks {\it completely} into the FQH droplet, with the ``smile" reducing  to a $U(1)$ vortex line on the pancake. That the resultant FQH droplet correctly carries the baryon charge  is assured by the gauge invariance of the Chern-Simons term (\ref{cs}). How this comes about can be explained in terms of a chiral bosonic edge mode~\cite{karasik1}.\footnote{This edge mode will be found to play a key role in accessing the EoS for massive compact stars~\cite{MR-PPNP}}  
In accordance with the global symmetries of QCD,   the $B=1$ baryon with $N_c=3$ quarks must then have spin $J=N_c/2=3/2$. This yields the high-spin baryon. Thus when the bag is shrunk by fiat to zero size, the Cheshire Cat smile will go into the vortex line in the FQH droplet. For instance for $N_c=3$, this picture yields the $\Delta (3/2,3/2)$.  The same  $\Delta (3/2,3/2)$ also appears in the rotational quantization of the skyrmion with $N_f=2$ which comes from the $\infty$-hotel mechanism which does not work for the $N_f=1$ baryons. These two descriptions present an aspect of the dichotomy problem: Whether or how they are related?
\subsubsection{Baryons for  $N_f\geq 2$}\label{dp}
Instead of a flavor-singlet quark, now consider the jail-breaking scenario of the doublet  u  snd  d quarks. There seems to be nothing to forbid the quark from flowing, instead of dropping into the infinite hotel giving rise to a skyrmion, into the $y$ direction as the flavor-singlet quark did to compensate the anomaly generated by the bag wall. Or is there? This is the question raised. 

Now let us blindly apply the same anomaly-flow argument in  CCP  to the $N_f$-flavored quark. The spin-flavor symmetry for the flavor $N_f\neq 1$ will of course be different. Given $N_f=2 $,  we expect to have a non-abelian Chern-Simons field $\mathbf A_\mu$ in place of the abelian $a_\mu$~\cite{MNRZ},
\be
\frac {N_c}{4\pi}\int_{2+1} {\rm Tr}\left(\mathbb A d\mathbb A+\frac 23 \mathbb A^3\right).
\label{nonabliean}
\ee
This presents an alternative jail-break scenario to the infinite-hotel one. 

But there arises the question: What makes nuclear matter (at $n\sim n_0$) realized as a state of skyrmions as Nature seems to indicate, instead of stacks of fractional quantum Hall pancakes or pitas~\cite{karasik1} or  combinations of the two? Is the non-abelian Chern-Simons droplet a meta-stable state absent at low density but could figure at high density? This question is addressed below  following the recent developments on the role of two hidden symmetries,  flavor local and scale,  intervening at high density involving fractional quantum Hall droplets~\cite{karasik1,karasik2,kitano}.

\subsection{Fermion Number and Hall Conductivity\\ on Domain Wall}\label{domainwall}
The Cheshire Cat Principle posits that physics should not depend on confinement size. It could be more appropriately phrased even as a  gauge dependence in gauge theories~\cite{CC-gauge}. This CCP on the confinement size  was proven by showing that the baryon charge does not depend on  $R$~\cite{CCP}. This follows from that  the baryon charge is topological in any dimension.    In (3+1)D, however, there is no exact bosonization and hence there is no exact CCP for other than the baryon charge, although approximate CCP holds for certain quantities like the flavor singlet axial charge of the proton~\cite{FSAC}.  
In \cite{MNRZ}, the CCP was established also for the $N_f=1$ baryon for the baryon charge with the fractional quantum Hall droplet replacing the skyrmion  for $N_f\geq 2$.\footnote{In \cite{MNRZ}, the bag boundary is taken as a domain wall. Whether the bag boundary can indeed be thought in terms of a domain wall is not clear and remains to be examined in detail. In modern developments in gauge theories,  the concept of domain wall (together with ``interface") plays a singularly important role. This is particularly so, in particular in QCD, and is a huge subject in the literature.  We insert a (much too) brief comment as a footnote just to give an idea. The $\theta$ dependence in QCD with massless quarks makes the pertinent case in this note as will be elaborated below.

As well known, the CP symmetry is  spontaneously broken for the vacuum angle $\theta=\pi$. Suppose  $\theta$ varies from 0 to $2\pi$.  There results a domain wall with Chern-Simons theory on it. Now when quarks are massless, since the bulk property of the theory depends on $m^{N_f} e^{i\theta}$,  the $\theta$ dependence is eliminated, replaced by a shift of $\eta^\prime$.  This is the anomaly cancelation restoring  CP symmetry in 4D. Thus the emergence of $\eta^\prime$ in the problem.}  

In oder to understand what's going on, let us re-derive the CCP result for $N_f=1$ baryon of \cite{MNRZ} in (3+1)D by considering the bag boundary as an extremely thin ``domain wall" located at $x_3=0$. Following \cite{vassi},  we will consider quantized Dirac fermions -- say, ``quarks" -- in interaction with a background $U(1)$ gauge field $a_\mu$, and scalar $\sigma$ and pseudo-scalar $\pi$ fields
\be
\cal{L}=\bar{\psi}{\cal  D}\psi\label{vassi-L}
\ee
with 
\be
{\cal D}=i\gamma^\mu (\del_\mu-ig a_\mu)-(\sigma+i\gamma_5 \pi), \ \sigma^2 +\pi^2=1.
\ee
$a_\mu$,  the $U(1)$ component of HLS,   will be more precisely specified below.

Consider the background fields changing rapidly near $x_3=0$ and going to asymptotic values. 
One is interested in the {\it vacuum} baryon number $B$ given by 
\be 
B=-\frac 12\eta(0,H)
\ee
where  $\eta (s, H)$ is the  spectral asymmetry that was computed in \cite{GJ} (for the infinite --hotel scenario)
\be
\eta (s, H)=\sum_{\lambda >0} \lambda^{-s} -\sum_{\lambda<0} (-\lambda)^{-s}\label{SA}
\ee
where $\lambda$ is the eigenvalues of the Dirac Hamiltonian $H$.  
With some reasonable approximations, it was obtained in \cite{vassi} that
\be
B=-\frac{g}{4\pi^2} \theta|^{x^3=+\infty}_{x^3=-\infty}\int d^2x f_{12}
\ee
where $\theta \equiv ({\rm arctan}(\pi/\sigma))$ and $f_{\mu\nu}$ is the gauge field tensor. Note that the vacuum fermion number $B$ has two components, first the Goldstone-Wilczek fractionalized fermion number~\cite{goldstone-wilczek} and the other the magnetic flux through the $(x,y)$ 2-d plane  

Consider next a domain wall background defined by the fields $\sigma$ and $\pi$  that depend on $x^3$ only. The one-loop effective action in the non-static background is found to give the parity-odd action\footnote{Why the parity-odd action becomes relevant is explained below.}
\be
S=\epsilon^{\mu\nu\rho 3} \int d^4x d^4y G(x,y)a_\mu (x)\del_\nu^y a_\rho (y) 
\ee
where $G$ is a complicated non-local function of $x^3$, $y^3$ and $z^\alpha=x^\alpha-y^\alpha$, $\alpha=0,1,2$. In the long-wavelength limit in the form factor $G$,
the action can be written as a Chern-Simons term 
\be
S=g^2 \frac{k}{4\pi} \epsilon^{\mu\nu\rho 3} \int d^3y^\alpha a_\mu(y^\alpha,0)\del_\nu a_\rho (y^\alpha,0)\label{S}
\ee
with
\be
g^2\frac{k}{4\pi} = \int d^3x^\alpha dy^3 dx^3 G(z^\alpha, x^3,y^3).\label{k}
\ee
Here $k$ can be identified as the ``level" in the level-rank duality of the Chern-Simons term. 

At this point one can make contact with what was done in the CCP structure~\cite{MNRZ}. For this consider the domain wall located at $x_3=0$ with the ``quark" modes inside the bag $x_3 < 0$ corresponding to  the Cheshire Cat smile.  The $U(1)$ field in (\ref{vassi-L}) could be considered,  as suggested in \cite{karasik1,kitano},  to be the $\omega$ field when the vector mesons $\rho$ and $\omega$ in HLS are treated as the color-flavor locked $U(N_f)$ gauge fields dual to the gluon fields in QCD~\cite{KKYY}. Then the  $\omega$ field can be taken as the Chern-Simons field that captures \`a la CCP the strongly-correlated excitations outside the bag. Now for $U(N_f)_{-N_c}$ dual to $SU(N_c)_{N_f}$ spontaneously broken, the vortex configurations in three dimensions made up of $\rho$ and $\omega$ carry magnetic and electric charges of $U(1)^{N_f}$. The electric charge in the CS term can then be identified with  the baryon charge~\cite{kitano,KKYY}. This allows one to obtain  the vector current from the action $S$ (\ref{S}), the time component of which is
\be
J^0 (x)=\frac{1}{g}\frac{\delta}{\delta a_0(x)} S.
\ee
The baryon number is~\cite{vassi} 
\be
B=\int d^3x J_0 (x)=\frac{gk}{2\pi}\int f_{12} d^2x.
\ee
Setting the Dirac quantization for the magnetic flux threading the vortex~\cite{KKYY}
\be
\frac{g}{2\pi}\int  f_{12} d^2x =1
\ee
one finds the baryon charge equal to the level
\be 
B=k.
\ee 
This is the baryon charge lodged {\it in the vacuum.}

Now to make the connection \`a la \cite{vassi} to the Cheshire Cat scenario discussed in \cite{MNRZ}, we identify the chiral angle for  $\theta$ which is $ =\eta^\prime/f_{\eta^\prime}$ in \cite{MNRZ}, and impose at $x^3=0$ the Cheshire Cat boundary condition 
\be
(1-i\gamma^3 e^{i\gamma_5\theta})\psi|_{x^3=0}=0. \label{bc}
\ee
Then the change in baryon charge is given by
\be
\Delta B\approx \frac{\Delta \theta}{2\pi} 
\ee
where  $\Delta\theta$ is the jump of the $\eta^\prime$ field across the chiral bag boundary. This is the same result obtained in \cite{MNRZ}. The Cheshire Cat dictates the baryon charge $B_{out}=1-B_{in}$ to be lodged in the Chern-Simons action
\be
S^\prime =g^2 \frac{k^\prime}{4\pi} \epsilon^{\mu\nu\rho 3} \int d^3y^\alpha a_\mu(y^\alpha,0)\del_\nu a_\rho (y^\alpha,0)\label{S'}
\ee
so it must be that
\be
k^\prime=1-k.
\ee

Here are two important points, among others, to note. First of all, as pointed out in \cite{vassi},  the Chern-Simons term (\ref{S}) or (\ref{S'}) by itself is  not topological. This is because the level $k$ or $k^\prime$ separately as defined is not an integer so the action is not gauge invariant, hence unphysical,  for $R\neq 0$ or $\infty$. The sum of the baryon charges of inside and outside is required by the anomaly cancellation. We believe this is related to the color anomaly found in the Cheshire-Cat in (3+1)D~\cite{colorleakage} explaining the tiny flavor-singlet axial charge $g_A^{(0)}\lsim 0.3$~\cite{FSAC}. 

Second one could have {\it naively} done the same analysis for the $N_f=2$ case with the pion fields included. That would have given rise to nonabelian CS theory with the same results as in the CCP strategy.  So one is back to the dichotomy problem.

\section{ The Dichotomy Problem}

\subsection{Indispensable Role of Vector Mesons}
We suggest that the key elements that provide the resolution of the dichotomy problem are the symmetries that led to the density-functional formalism G$n$EFT, namely, the hidden local symmetry and the scale/conformal symmetry. The degrees of freedom associated with these symmetries, the vector mesons and the dilaton, can be taken as emergent symmetries from strong nuclear correlations  ``dual" to QCD. 

To see how one arrives at this aspect, let us incorporate the $\eta^\prime$ field in the two-flavor chiral field in HLS Lagrangian as
\be
U=\xi^2=e^{i\eta^\prime/f_\eta}e^{i\tau_a\pi_a}.
\ee
The crucial observation made by Karasik~\cite{karasik1,karasik2} is that what is called  ``hidden" Wess-Zumino term in the HLS Lagrangian\footnote{This term corresponds to the ``homogeneous" Wess-Zumino term of \cite{HY:PR} with its (arbitrary) four coefficients  constrained by, e.g., the vector dominance (VD). We refer to both as ``hWZ term."} unifies the baryon currents for both the FQH droplet and the skyrmions~\cite{dichotomy}.  It was noted that in the effective field theory that contains both $\eta^\prime$ and the HLS fields, the $\eta^\prime$ cusp that accounts for the jump from one vacuum to the other at $\eta^\prime=\pi$ does not appear. Thus the effective theory containing the HLS fields in the presence of $\eta^\prime$ captures the emergent theory on the $\eta^\prime$ domain wall.  This suggests that it is more efficient and simpler to resort to the bosonic Lagrangian from which both the skyrmions and the FQH droplets emerge as solitons. It is not clear how to bring G$n$EFT to the density regime involved which could be higher than what's relevant in compact stars. This is an open problem.\footnote{It seems feasible to formulate this problem via the nonlinear bosonization of the Fermi surface giving rise to soft modes of hidden symmetries~\cite{coadjoint-orbit}.}

Now let's consider tweaking the baryonic matter by increasing density in $\chi$HLS Lagrangian where the hWZ terms figure. The density for compact stars is $\lsim 10 n_0$. The relevant structure is more or less captured by the approach to the dilaton limit fixed point  -- Section \ref{DLFP} -- and the emergent pseudo-conformal symmetry (PCS) -- Section \ref{EPCS} -- which say that $g_A^{DL}\to 1$ and $f_\pi\to f_\chi\sim m_0$ in the range of density involved in massive stars. As argued in \cite{dichotomy}, the FQH droplet should become relevant at some high density at which $f_\chi\to 0$. This density must then be (much) higher than that reached at the DLFP close to the GD's IR fixed point. It is at this point the hWZ term exposes the Chern-Simons $\eta^\prime$ coupling term carrying the information on the FQH droplet with the correct baryons number. There the $\omega$ field in HLS Lagrangian can be identified as the Chern-Simons field.

In \cite{kitano}, a scenario different from that of \cite{karasik1,karasik2} is suggested for the role of hidden local symmetry. There the coupling of the Chern-Simons fields in the bulk couple with the edge modes of vector mesons making the vector mesons gauge bosons. At the moment which scenario is preferred is not clear. But what's absolutely clear is that hidden symmetries ``dual" to QCD symmetries (e.g., HLS vector mesons as Seiberg-dual to the gluons~\cite{KKYY}) must be essential for the correct description of the phase structure at high density. One cannot say whether the compact star density reaches the appropriate density beyond the PC regime. If future refined gravity wave observations were to indicate significant deviations from the PC sound speed predicted in our approach, this would give a hint to the possible role of the FQH droplets. 

An interesting observation here is that the Chern-Simons field coupling to the FQH droplet in the hWZ term
\be
{\cal L}_{\rm CS\eta^\prime}=-\zeta \frac{N_c}{4\pi} J_{\mu\nu\alpha}\omega^\mu\del^\nu \omega^\alpha
\ee
with the topological $U(1)$ 2-form symmetry current
\be
J_{\mu\nu\alpha}=\frac{1}{2\pi}\epsilon_{\mu\nu\alpha}\del^\beta \eta^\prime
\ee
requires $\zeta=1$ to have gauge invariance\footnote{In the GD scheme~\cite{GDS} we are adopting, in the absence of the $\eta^\prime$ field, each of the four hWZ terms $a=1,2,3,4$ can have a factor $\Big(c_a + (1-c_a)\big(\frac{\chi}{f_\chi}\big)^{\beta^\prime}\Big)$ with $c_a$ an unknown constant and $\beta^\prime$ the anomalous dimension of the gluon stress tensor that multiplies the scale-invariant term.}. This means that the ``scale-symmetry breaking constant" $c_{hWZ}$ in front of the hWZ term is $\zeta=c_{hWZ}=1$ whereas in the absence of the FQH droplets  it could be that  $\zeta=c_{hWZ} << 1$  at scale-chiral symmetry  restoration with $f_\chi=f_\pi=0$~\cite{omega-ma-rho}.

\subsection{Dense Matter as ``Sheets" of Pancakes/Pitas}

As noted, at low density, the $N_f=2$ quarks in the bag must be tending to fall into the infinite hotel, hence giving rise to skyrmions in (3+1)D.  This may be ``driven" by the parameters of the Lagrangian that unifies the $N_f\geq 2$ and $N_f=1$ baryons  to have the $B^{(0)}$ effect  {\it suppressed} at low density.  However as density increases,   the parameter change in the baryonic scale-symmetric Lagrangian ${\cal L}_{\psi\chi {\rm HLS}}$ in G$n$EFT that distorts the baryon current from the unified current to the ${N_f=1}$ current could transform the EoS state toward the Chern-Simons QFT structure. One possible scenario for this is indicated in the recent skyrmion crystal analyses of dense matter where an inhomogeneous structure is found to be energetically favored ultimately over the homogeneous one at high density.  It has been found that dense matter consists of a layer of sheets of ``lasagne" configuration with each sheet supporting half-skyrmions~\cite{PPV}.
%
The constituents of this layer structure are quasi-fermions consisting of fractionalized quasiparticles  of 1/2 baryon charge, possibly deconfined as conjectured below, appearing in baryon-quark continuity at a density $\sim n_{1/2}$, drastically different from those of the pasta structure discussed for the dilute outer layer of compact stars. In the Skyrme model (with pion field only) used  in \cite{PPV}  the quartic (Skyrme) term effectively encodes massive degrees of freedom,  including the hidden local fields, the monopole structure hidden in half-skyrmions etc. described above.  It appears feasible to formulate this ``sheet dynamics"  by a stack of FQH pancakes or pitas with tunneling half-skyrmions between the stacks, somewhat like arriving at the Chern-Simons field theory structure of FQHE in (2+1)D with a stack of (1+1)D quantum wires~\cite{quantumwires}\footnote{There are many papers on this matter in the literature. A good article with many relevant references is  \cite{quantumwires}}. In the tunneling process it may be possible that the half-skyrmions transform to 1/3-charged quasiparticles resembling  quasi-quarks as discussed in \cite{ vento-half-skyrmion}. On the domain wall, the 1/3-charged objects could behave as ``deconfined" quarks as discussed in \cite{deconfinedDW}. This could then explain why the composite quasi-fermions given in G$n$EFT behave similarly to the ``deconfined quarks" in the core of massive compact stars~\cite{deconfined}. 
\section{Conclusion}
The question addressed in this note was: Is it feasible with a single unique effective Lagrangian to address the equation of state from normal nuclear matter to massive compact-star matter resorting to only one set of degrees of freedom with the vacuum sliding with density but without  phase transitions ? Put in another way, how far can one go with such a ``unified" formalism without getting into fatal conflict with either empirical or theoretical constraints?

Influenced by strikingly successful developments in strongly correlated condensed matter physics  together with impact on particle physics, an extremely simplified approach to the EoS for massive compact stars is formulated in terms of topology change to account for possible ``continuity" from seemingly hadronic variables to QCD variables at high density $n_{1/2}\gsim 3n_0$. The single Lagrangian adopted in this study consists entirely of hadronic variables, the pion and the nucleon that figure in the standard nuclear EFT plus the massive degrees of freedom $\rho$, $\omega$ and $\chi$ associated with hidden local and scale symmetries. The role of the topology change is to endow what could be identified as Kohn-Sham-type ``density-functional" structure in the parameters of the effective Lagrangian that are supposed to capture the topological structure of QCD variables in dense medium. We further extended the scenario with the possible intervention of the FQH droplets $B^{(0)}$s going beyond the DLFP, brining in non-Fermi baryonic matter with scale-chiral restoration.

In this approach, there are neither explicit quark degrees of freedom nor  strangeness flavor as  in the standard approaches~\cite{baymetal,alford} and in other variations with bag models~\cite{MIT}.  It is  possible of course that  there be  corrections  to the approximations made --  given the admittedly drastic oversimplification  -- that could,  quantitatively though not qualitatively, modify the results. There is however one serious potential obstruction to G$n$EFT.  Should future measurements map out precisely the behavior of the sound velocity in the range of density $3\lsim n/n_0\lsim 7$ and falsify the precocious onset of,  and the convergence to,  the PC sound velocity, then that would bring a serious obstruction to the notion of the emergent symmetries, particularly hidden scale symmetry distinctive of the theory. That would then ``torpedo"  the G$n$EFT.   If however it is not ``torpedoed,"  then our approach with the encoded  ``duality" to QCD in approaching the chiral phase transition as well as confinement as argued recently by string-theory-inclined theorists~\cite{komargodski,karasik1,karasik2,kitano,KKYY}  will bring a totally new perspective to nuclear physics, a paradigm almost totally foreign to nuclear theories. 

A most interesting future direction would be to map the ``generalized" sheet structure of Chern-Simons QFT in the topological sector conjectured above to an improved G$n$EFT phrased in Wilsonian-type Fermi-liquid theory more powerful and realistic than what has been achieved so far, perhaps along the line of the nonlinear bosonization approach accessing non-Fermi liquid state. It would offer a clear  resolution of the dichotomy problem and escape the possible obstruction to the G$n$EFT.

\subsection*{Acknowledgments}
We are grateful for collaborations and/or  discussions with Tom Kuo,  Hyun Kyu Lee, Yong-Liang Ma, Maciej Nowak, Won-Gi Paeng,  Sang-Jin Sin and Ismail Zahed during and after the WCU/Hanyang Project.

\end{document}